\documentstyle[11pt,epsfig]{article}
\newcommand{\Li}[2]{{\mbox{Li}}_{#1}\left(#2\right)}
\begin{document}

\begin{flushright}
 {MZ-TH/97-37}\\
 {Freiburg-THEP 97/29}\\
 {hep-ph/9712205}\\
 {November 1997}
\end{flushright}

\vspace{2 cm}

\begin{center}
 {\bf \Large
${\cal O}(\alpha_s^2)$ corrections to $b \to c$ decay at zero recoil
 }\\[3ex]
J. Franzkowski$^{(1)}$ and J.B. Tausk$^{(2)}$
\\[3ex]

(1) Institut f\"ur Physik,
Johannes Gutenberg-Universit\"at,\\
Staudinger Weg 7, D-55099 Mainz, Germany
\\[1ex]

(2) Fakult{\"a}t f{\"u}r Physik,
Albert-Ludwigs-Universit{\"a}t Freiburg,\\
Hermann-Herder-Stra{\ss}e 3,
D-79104 Freiburg, Germany

\vspace{2 cm}

\begin{abstract}
Analytic formulae are presented for the two-loop perturbative QCD
corrections to $b \to c$ decay at the zero recoil point, which
are required for the extraction of $|V_{bc}|$ from measurements of
exclusive $B\to D^* l \nu$ decays. The results are in agreement with
those in \cite{Czarnecki,CM}. Some comments on the numerical evaluation
of the diagrams involved are made.
\end{abstract}

\end{center}

\section{Introduction}
The magnitude of the Kobayashi-Maskawa matrix element $V_{cb}$ can be
determined experimentally by observing the decays of $B$ mesons
produced in $e^+e^-$ collisions. In particular, one method requires
the measurement of the rate of the exclusive semileptonic decay
$B \to D^* l \nu$ (see, eg., \cite{RichmanBurchat}). From the
decay rate, measured at the zero recoil point, i.e., in the special
kinematical situation where the $D^*$ is produced at rest in the
$B$ rest frame, one can then extract the value of $|V_{cb}|$. This
method has been used by experiments where $B$ mesons are produced
on the $\Upsilon(4S)$ resonance \cite{Y4S} and on the $Z$ resonance
\cite{LEP}. The statistical and systematic errors are currently of
the order of $5\%$. However, in the future, with the CESR collider
at Cornell running at increased luminosity, the asymmetric
$B$ factories at KEK and SLAC coming into operation, and further $B$-physics
experiments to be conducted at the hadron accelerators Tevatron,
HERA and LHC (see, e.g., \cite{Stone}),
the errors are expected to come down to around $1\%$ or less \cite{Kurimoto}.
Thus, it is important that the theoretical input, that is needed to
extract $|V_{cb}|$ from the measured decay rates, be known with equal
precision. The purpose of this paper is to present one component of
that theoretical input, namely the second order perturbative QCD
corrections to the decay of a $b$ quark into a $c$ quark at
zero recoil. The other part of the theoretical input consists of
non-perturbative corrections which are described by an expansion in
the heavy quark masses $m_b$ and $m_c$. For a review of
heavy quark theory and further references, see, e.g., \cite{BSU}.

The reason for using the zero recoil point is that at that point, the
non-perturbative contributions are suppressed by a factor of
$\Lambda_{QCD}^2/m_c^2$, because of an additional symmetry that exists
in the infinite quark mass limit. At the same time, however,
the zero recoil condition simplifies the kinematics of the decay
$b \to c W$ to such an extent, that a complete analytical calculation
of the perturbative corrections at the two-loop level becomes feasible.

\begin{figure}
\begin{center}
\Large
\[
\begin{array}{ccc}
\epsfig{file=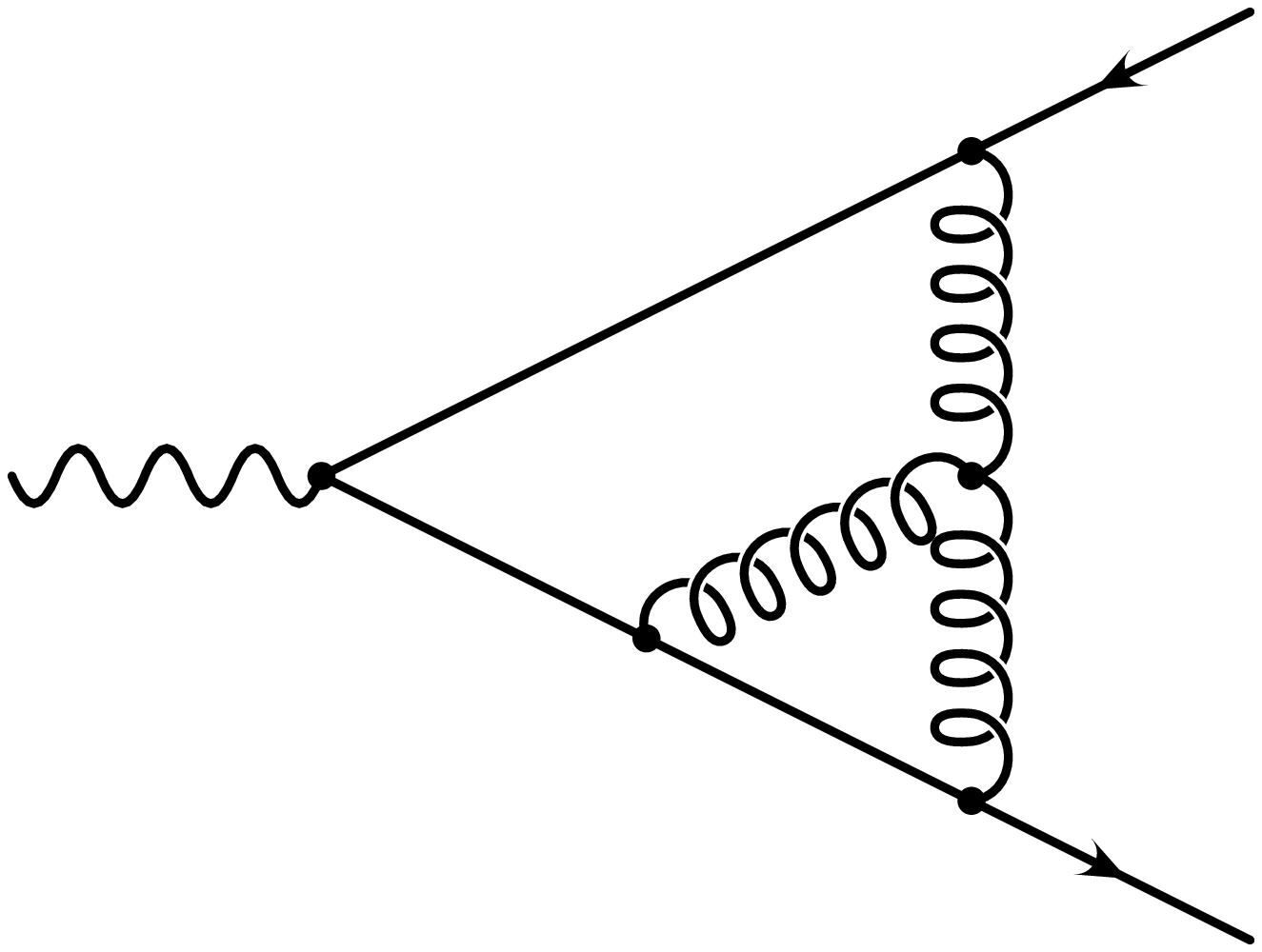,width=4cm}
&
\epsfig{file=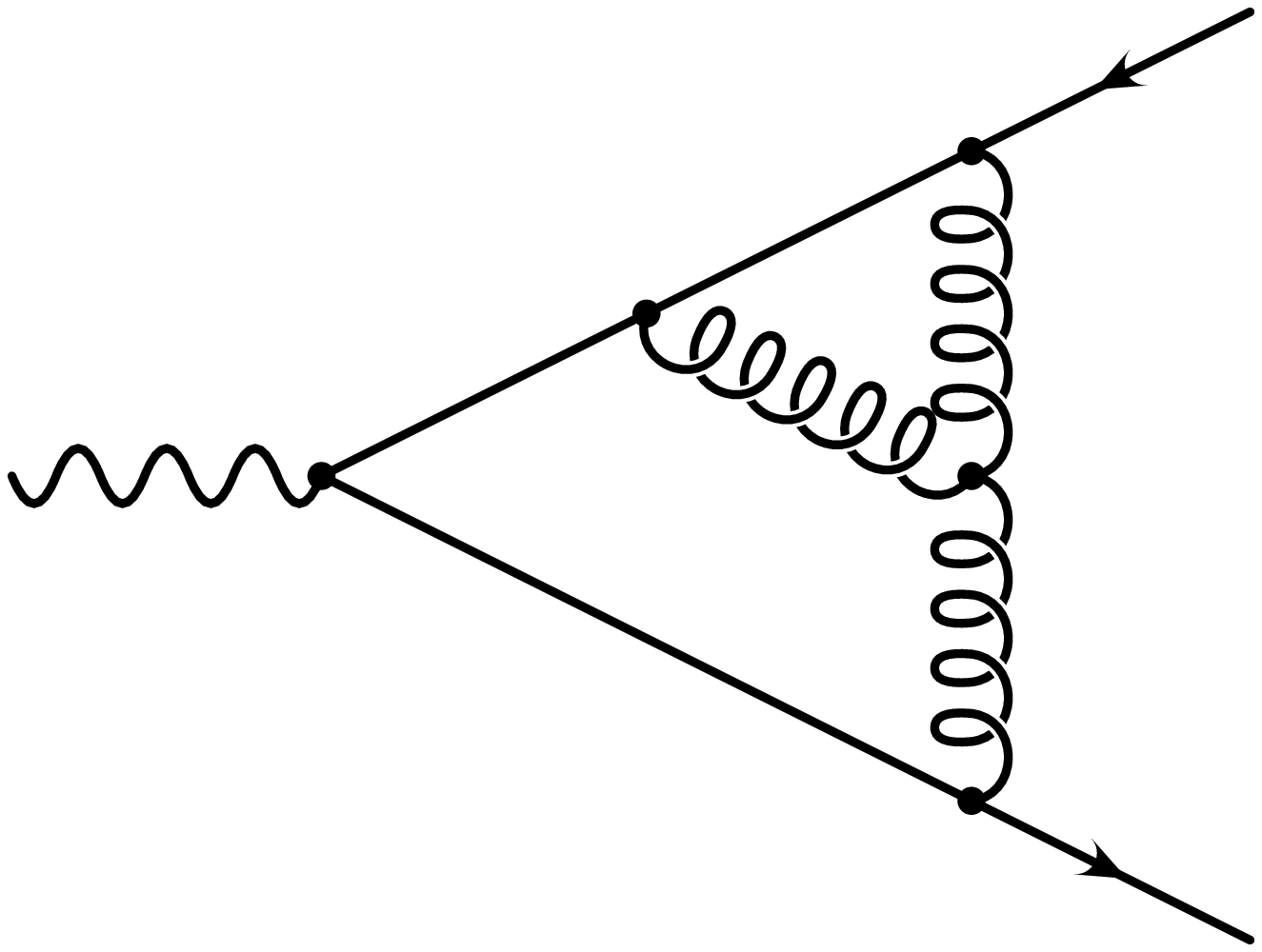,width=4cm}
&
\\
a_1&a_2&
\\[3mm]
\epsfig{file=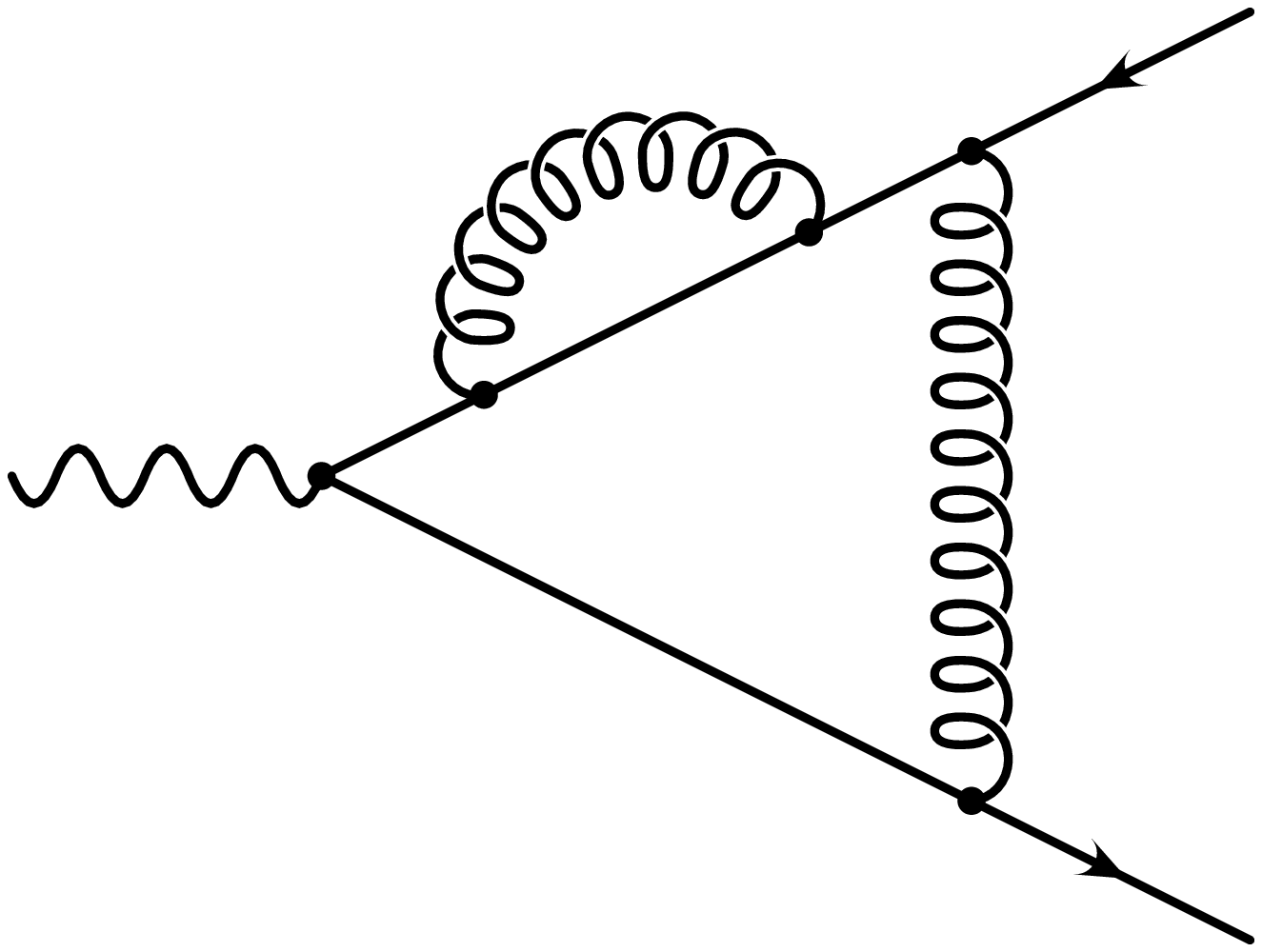,width=4cm}
&
\epsfig{file=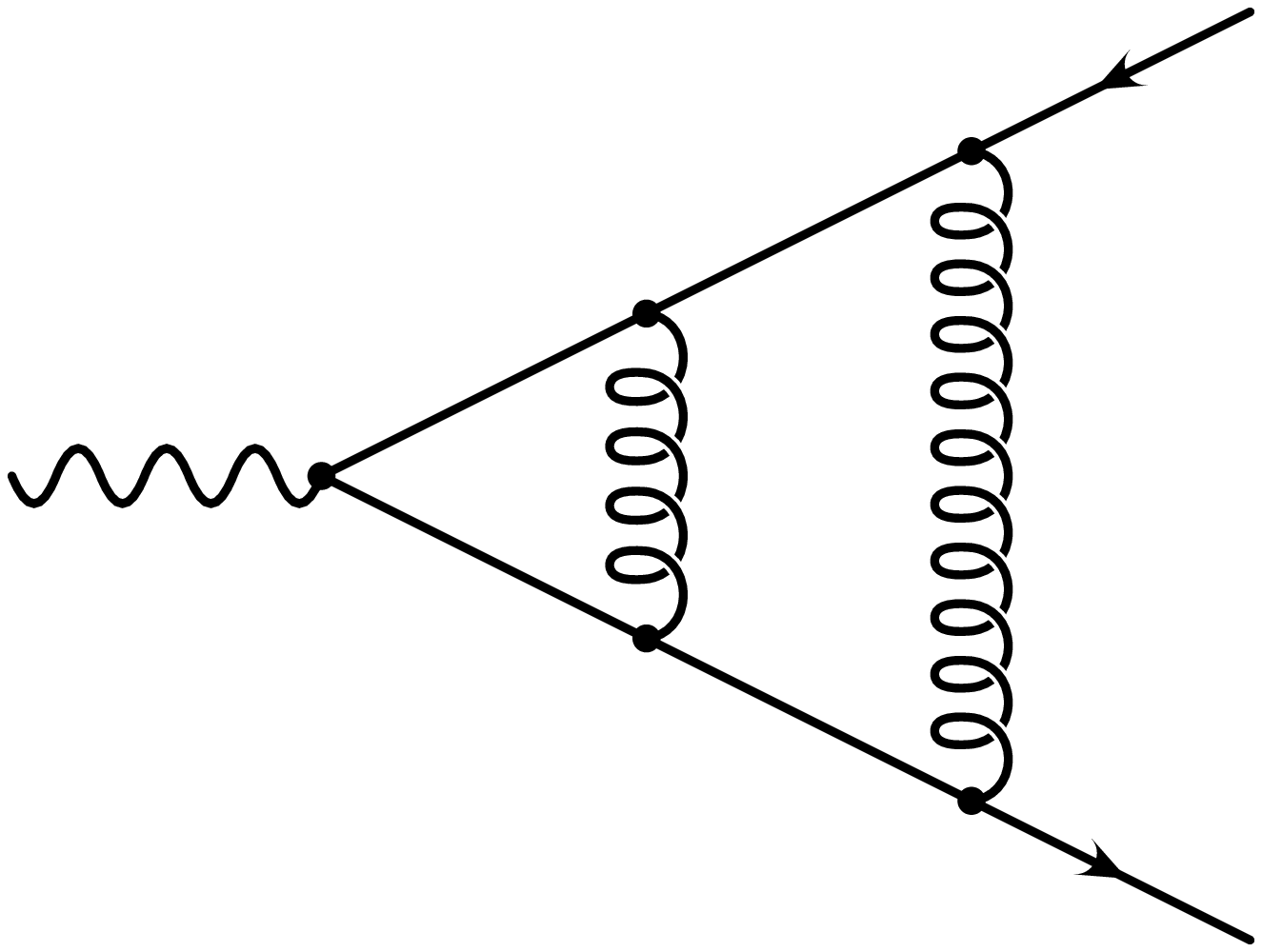,width=4cm}
&
\epsfig{file=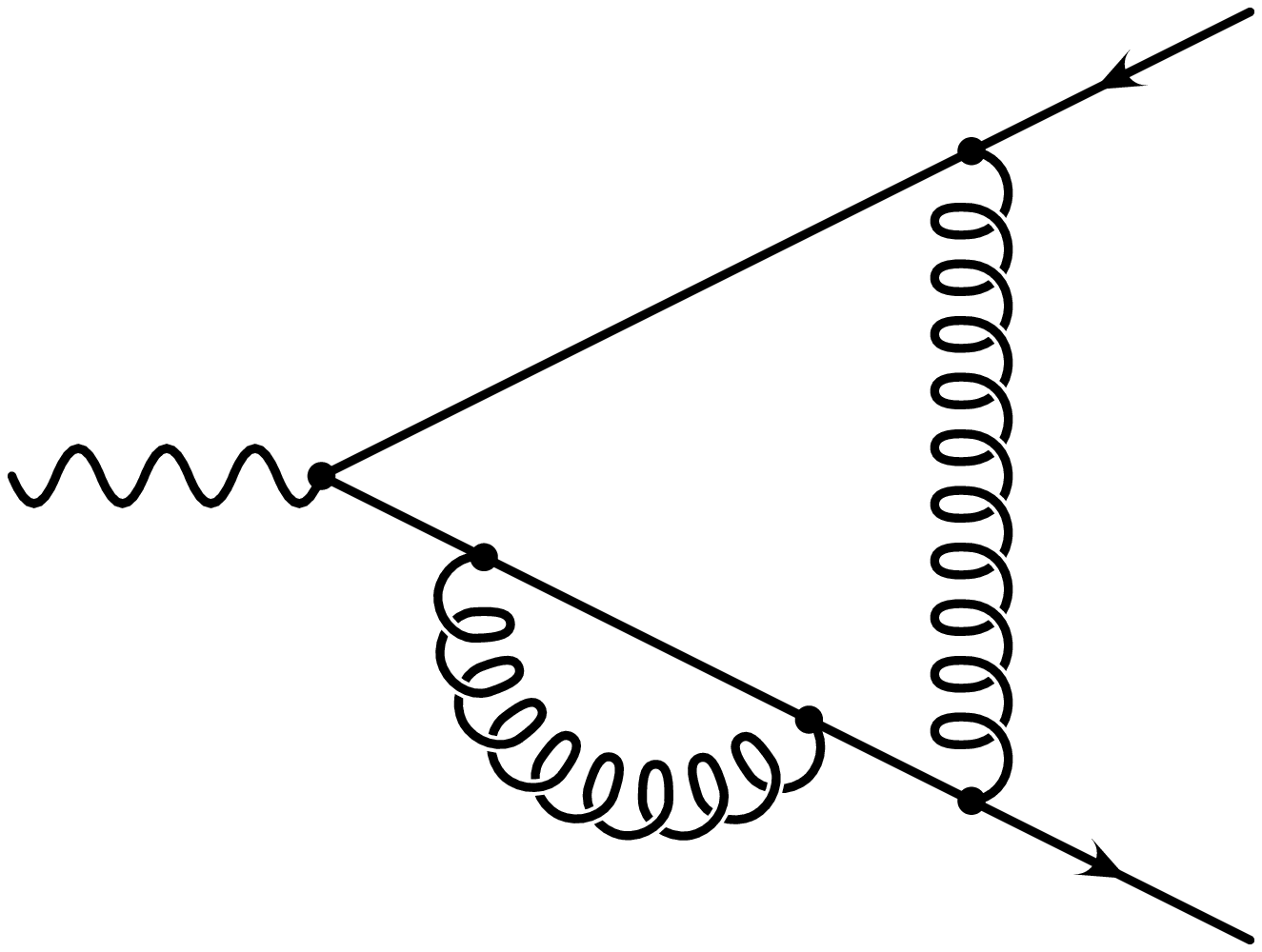,width=4cm}
\\
b_1&b_2&b_3
\\[3mm]
\epsfig{file=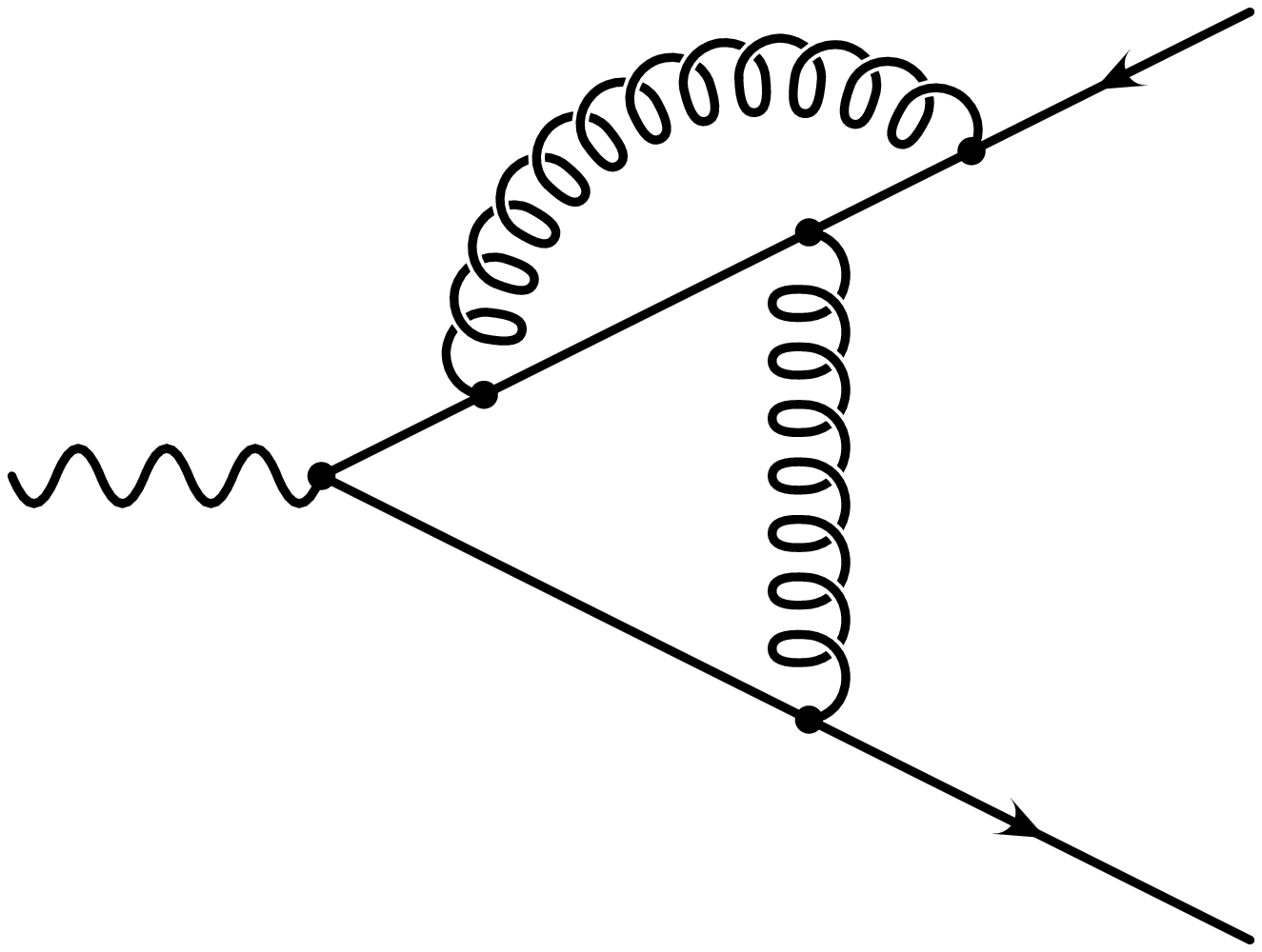,width=4cm}
&
\epsfig{file=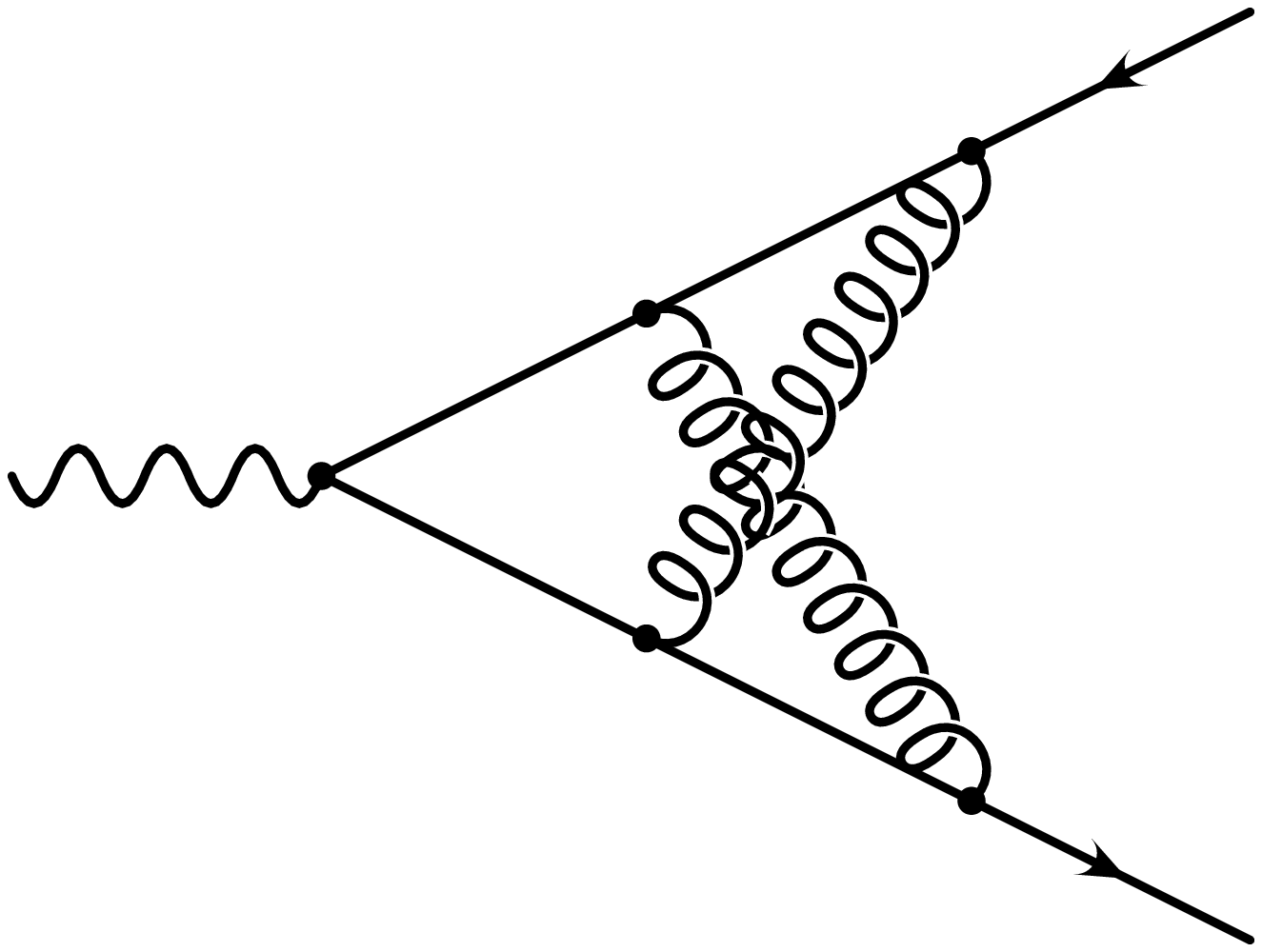,width=4cm}
&
\epsfig{file=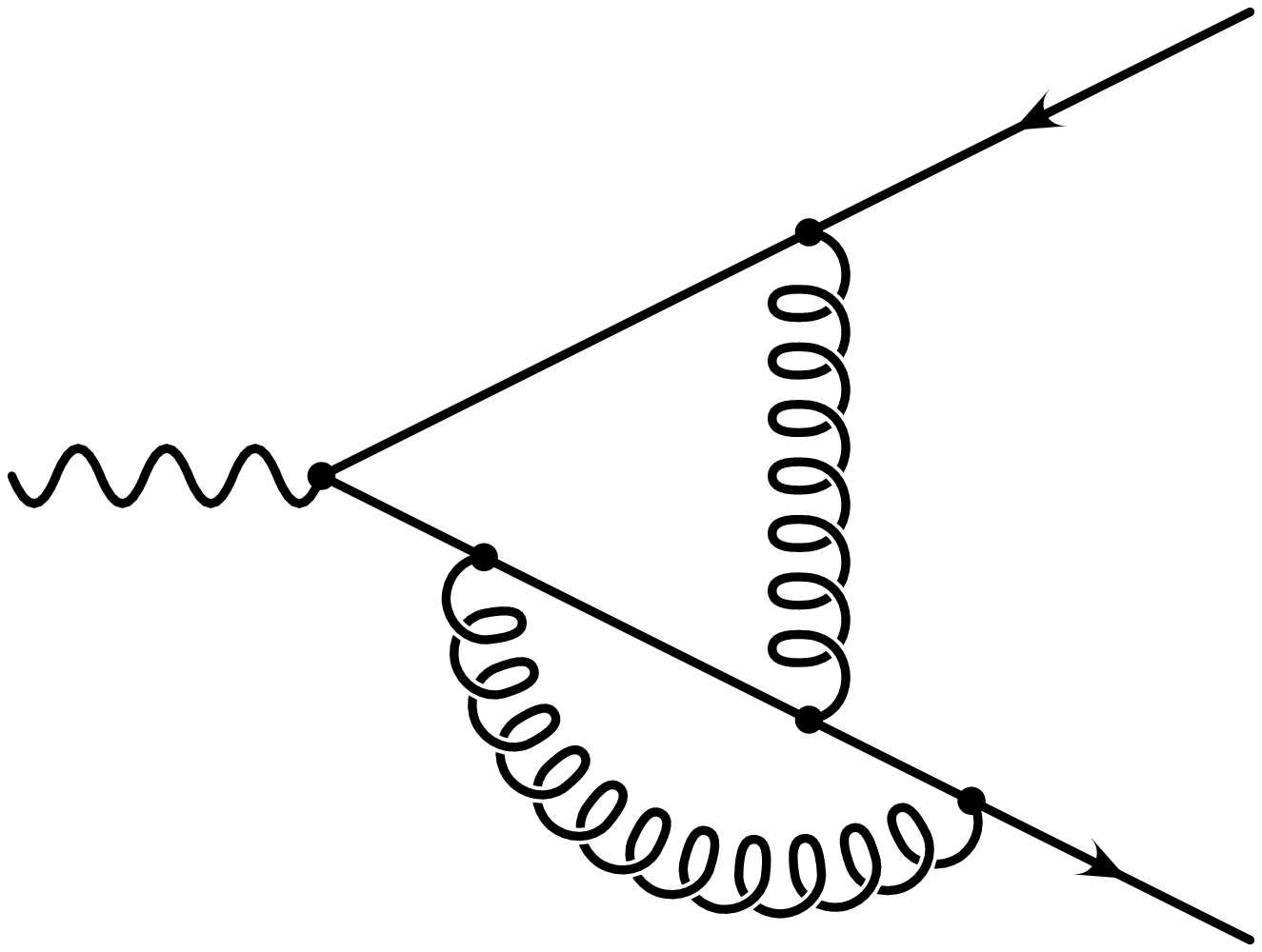,width=4cm}
\\
c_1&c_2&c_3
\\[3mm]
\epsfig{file=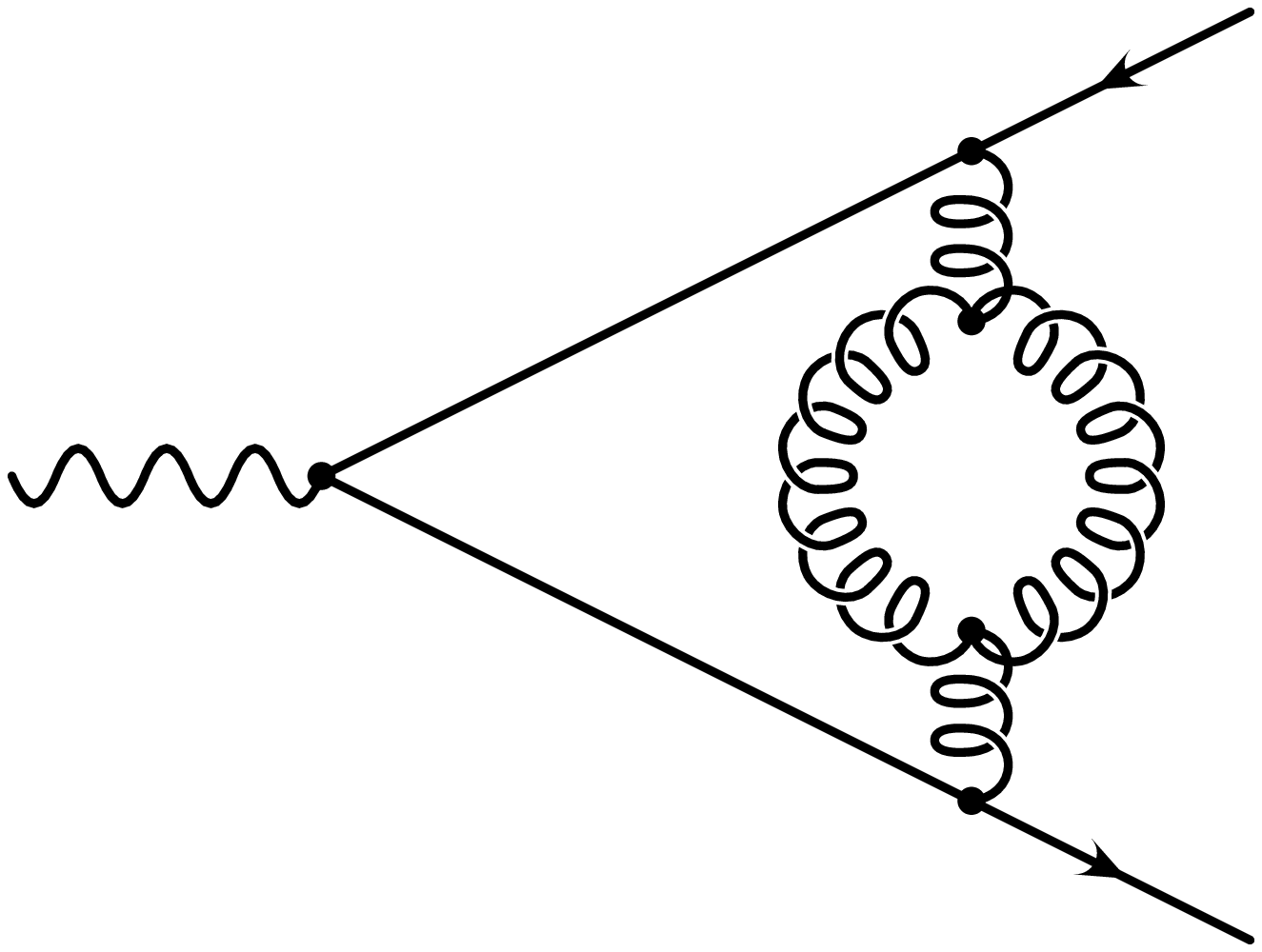,width=4cm}
&
\epsfig{file=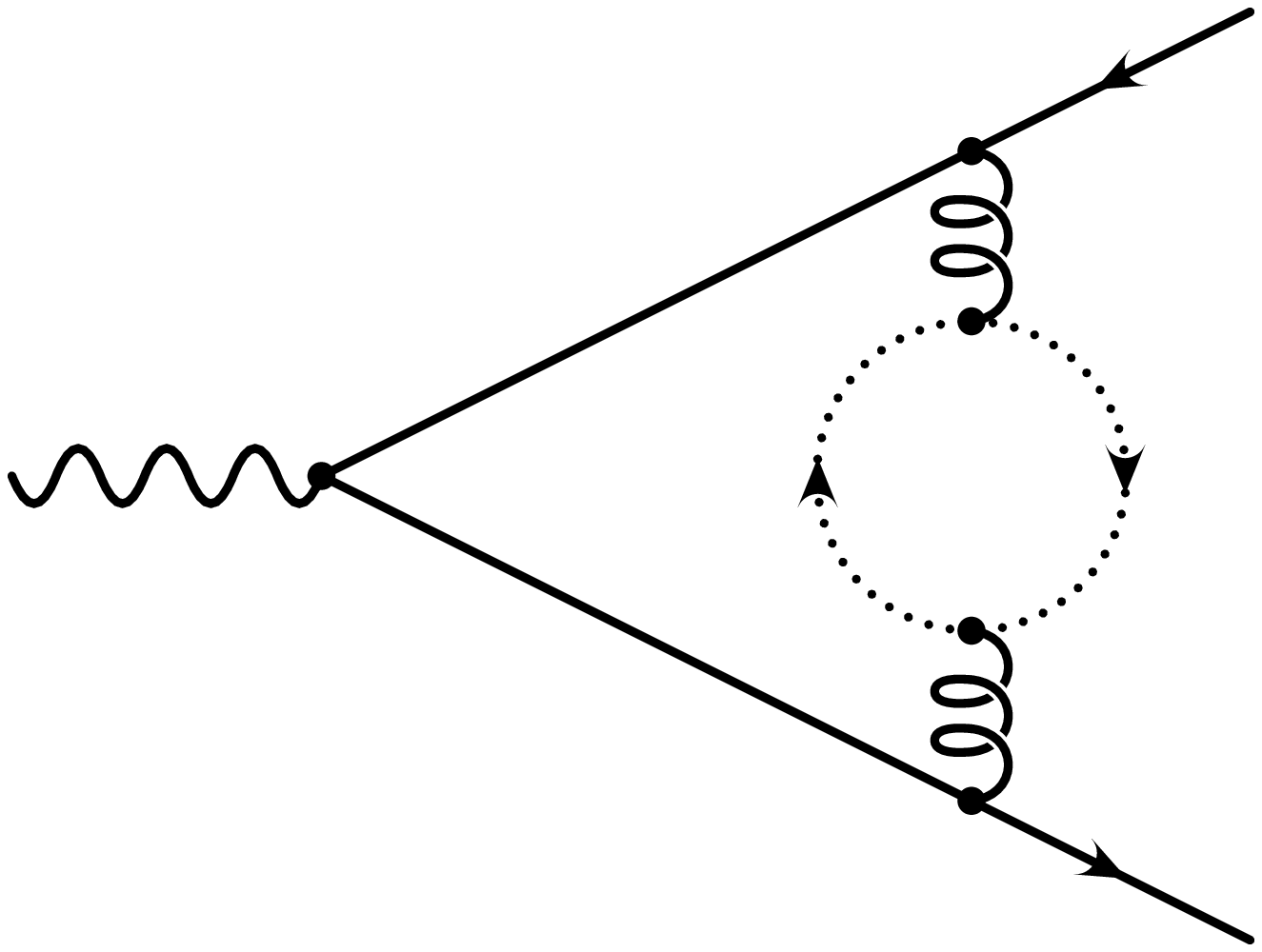,width=4cm}
&
\epsfig{file=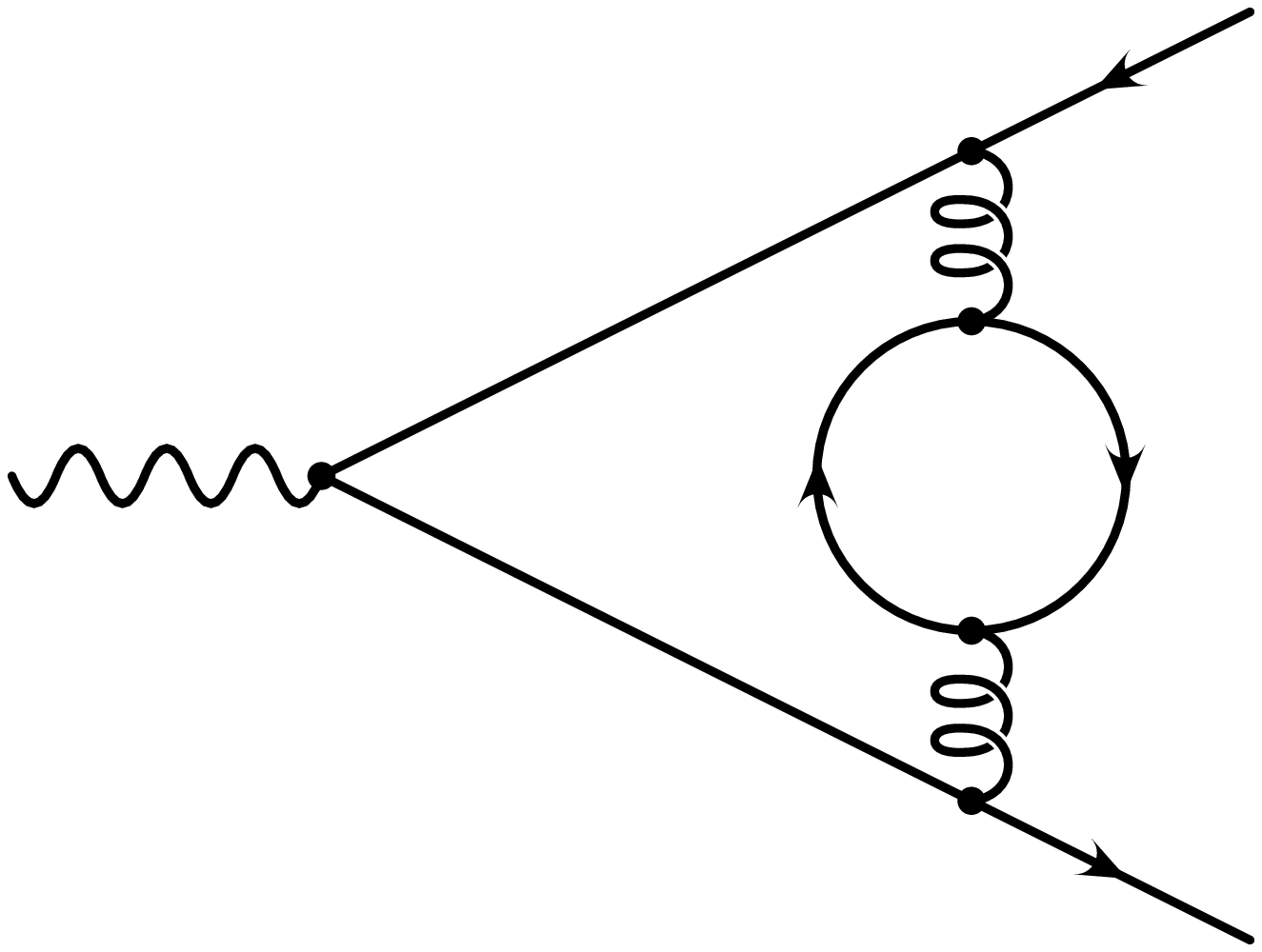,width=4cm}
\\
d&e&f
\end{array}
\]
\normalsize
\end{center}
\caption[]{Irreducible Feynman diagrams contributing to $b\to c W$ at
           order $\alpha_s^2$. The dotted line in diagram $e$ represents a
           Faddeev-Popov ghost. The fermion in the loop in diagram $f$
           can be either a light quark, a $b$ or a $c$ quark.}
\label{fig:diagrams}
\end{figure}

At tree level, the amplitude for $b \to c W$ is proportional to
$\bar{u}(c)\gamma^{\mu}\left(1-\gamma_5\right)u(b)$. In higher orders,
this gets modified into
$\bar{u}(c)\gamma^{\mu}\left(\eta_V - \eta_A\gamma_5\right)u(b)$,
where $\eta_{V,A}$ are given by perturbation series in $\alpha_s$.
(Only $\eta_A$ enters the expression for $B \to D^* l \nu$.)
At zero recoil, no other Lorentz structures appear. The order
$\alpha_s$ contributions were calculated in \cite{PaschalisGounaris}.
The Feynman diagrams that contribute to $\eta_{V,A}$ in order
$\alpha_s^2$ are shown in figure~\ref{fig:diagrams}. They were
calculated analytically in \cite{CM}, confirming the results
of a Taylor series expansion in $(m_b-m_c)/m_b$ \cite{Czarnecki} that
had been obtained earlier by one of the authors of \cite{CM}.
In this paper, we present an independent analytic calculation
of $\eta_{V,A}$. Although our results are expressed in a slightly
different way,
they are completely equivalent to the results of \cite{CM}.
Thus, we confirm the conclusions of \cite{Czarnecki,CM}.

Section~\ref{sec:calc} describes the main steps of the calculation. Although
the details are different, the methods we have used are nevertheless
related to
the ones employed and extensively discussed in \cite{CM}. The major part
of the work is the calculation of a set of scalar integrals, many of which
contain infrared divergences. One example is looked at more closely, and
various ways we checked our calculation are discussed. The final results
are presented in section~\ref{sec:res} and are followed by our conclusions
in section~\ref{sec:conc}. As a by-product of our work, we developed
a new version of a numerical technique for evaluating a class of two-loop
Feynman diagrams \cite{Kreimer2L2P}. The new version is much more suitable
for dealing with some of the rather special diagrams that occur in this
calculation. However, because the method itself is quite general and may be
useful for other problems as well, it is explained briefly in an appendix.
\section{Calculation of \boldmath{$\eta_{V,A}$}}
\label{sec:calc}
An important consequence of the zero recoil condition is that there is
no phase space available for gluon bremsstrahlung, $b \to c W g (g)$,
and thus, only the virtual corrections shown in figure~\ref{fig:diagrams}
are needed. We calculate them in the Feynman gauge and use dimensional
regularization for both ultraviolet and infrared divergences. We neglect
the masses of all the light quarks, and we leave out the diagram where
a $t$ quark loop is inserted into the gluon propagator. With these
restrictions, all diagrams can be written in terms of the following
nine propagator denominators:
\begin{equation}
\setlength{\arraycolsep}{0.05cm}
\begin{array}{rclrclrclrclrcl}
P_1 &=& (l+k).(2p_1+l+k)\,,\,\,
&
P_3 &=& l.(2p_1+l)\,,\,\,
&
P_5 &=& k.(2p_1+k)\,,\,\,
&
P_7 &=& k^2\,,\,\,
&
P_9 &=& (l+k)^2\,,\,\,
\\
P_2 &=& (l+k).(2p_2+l+k)\,,\,\,
&
P_4 &=& l.(2p_2+l)\,,\,\,
&
P_6 &=& k.(2p_2+k)\,,\,\,
&
P_8 &=& l^2\,,\,\,
\end{array}
\end{equation}
where $k$ and $l$ are loop momenta and $p_1$ and $p_2$ are the four-momenta
of the incoming $b$ quark and the outgoing $c$ quark, respectively.
Normally, seven of the denominators $P_i$ would be linearly independent,
but because of the zero recoil condition, which means that $p_1$ and $p_2$
are proportional to each other, $m_2 p_1 = m_1 p_2$, we have two
additional linear relationships between them:
\begin{equation}
\begin{array}{cc}
m_1 P_4 - m_2 P_3 = (m_1-m_2) P_8 , &
m_1 P_6 - m_2 P_5 = (m_1-m_2) P_7.
\end{array}
\end{equation}
This greatly simplifies the calculation.

After projecting the diagrams onto
the form factors $\eta_{V,A}$\footnote{
In the case of $\eta_A$, we use an anticommuting $\gamma_5$.}
and expressing all scalar products
of $k$, $l$, $p_1$ and $p_2$ that appear in the numerators in terms
of the $P_i$, which we do using Form~\cite{Form}, we obtain for each
one a combination of scalar integrals of the form:
\begin{equation}
\label{eq:generalscalarint}
\int \int \mbox{d}^d k  \; \mbox{d}^d l \;
\frac{1}{P_1^{\nu_1} P_2^{\nu_2} \ldots P_9^{\nu_9}} \, ,
\end{equation}
where $d=4-2 \epsilon$ (some of the $\nu_i$ can be negative). The
scalar integrals that occur can be classified into three groups.
The first group consists of integrals that can be factorized as
a product of two one-loop integrals, e.g.:
\begin{equation}
\int \int \mbox{d}^d k  \; \mbox{d}^d l \;
\frac{1}{P_3 P_6 P_7 P_8} \,
= \,   \int \mbox{d}^d l \; \frac{1}{P_3 P_8}
\times \int \mbox{d}^d k \; \frac{1}{P_6 P_7} \, .
\end{equation}

The second group consists of integrals in which either
$\nu_2=\nu_4=\nu_6=0$ or $\nu_1=\nu_3=\nu_5=0$, so that they only
depend on one mass scale. Such integrals also occur in on-shell
fermion self-energies and in anomalous magnetic moments, and
can be calculated using recurrence relations based on
integration by parts \cite{MN,BGS}. A detailed explanation of
the algorithm is given in \cite{GrayThesis}.

The third group contains the non-factorizable graphs that depend
on $m_1$ and on $m_2$. Although one can still derive relations
between them using integration by parts, those relations are
more complicated than in the case of just one mass scale. We
find them to be quite useful in a few cases, but are still
left with rather a large number of cases we have to calculate
from scratch.

Before actually calculating the remaining integrals in the third
group, it is worth while to investigate their analytic properties
by solving the Landau equations.
One finds that they can have singularities when
$m_1=0$ or $m_2=0$. At $m_1=m_2$ there are no singularities
because, although some poles in the propagators coincide at that point,
the integration contours are not pinched. However, there may
be singularities at $m_1=m_2$ on the analytic continuation to
higher Riemann surfaces. The following very simple one-loop
example illustrates these properties:
\begin{equation}
\label{eq:oneloopeg}
\int \mbox{d}^d k  \;
\frac{1}{k.(2p_1+k) \, k.(2p_2+k)}
= i \pi^{2-\epsilon} \, \Gamma(1+\epsilon)
\left( \frac{1}{\epsilon} + 2 - 2 \, \frac{m_1\log(m_1)-m_2\log(m_2)}{m_1-m_2}
\right)
\end{equation}
If we analytically continue the right hand side of this equation in,
say, $m_2$, going around the branch point at $m_2=0$ and then back to
$m_2=m_1$, $\log(m_2)$ changes into $\log(m_2) +  n (2 \pi i)$,
which no longer cancels $\log(m_1)$, and as a result, a pole appears
at $m_1-m_2=0$. In addition to the singularities just mentioned, a certain
subclass of the two-loop integrals (\ref{eq:generalscalarint}) can also
have singularities when they are analytically continued to the
point $m_1+m_2=0$. The integrals in this subclass correspond to
graphs that can be cut into two pieces by removing exactly
three massive quark propagators. They only occur in diagrams $c_1$, $c_2$,
and $c_3$ in figure~\ref{fig:diagrams}, and in diagram $f$, when the
quark in the loop is massive.
While this discussion of analytic continuations to negative masses
and higher Riemann surfaces may seem academic, keeping these properties
in mind while actually doing the integrations can be very helpful,
because it tells us which polylogarithms we can expect to occur in
the answer.

We have done most of the integrals needed using
Feynman parametric representations
of the kind described in \cite{RainerScharf}.
Sometimes, it is convenient to differentiate a diagram with respect
to one (or both) of the masses first, in order to reduce the
number of different kinds of propagators, and thus the number
of Feynman parameters, and then reintegrate with respect to the mass to
get the final answer. For the integration constant, we
can take the equal mass point, which belongs to the second
group discussed above. Another reason why we might want
to differentiate with respect to masses, is to make an
integral less infrared divergent.

\begin{figure}[htb]
\centerline{\epsfig{file=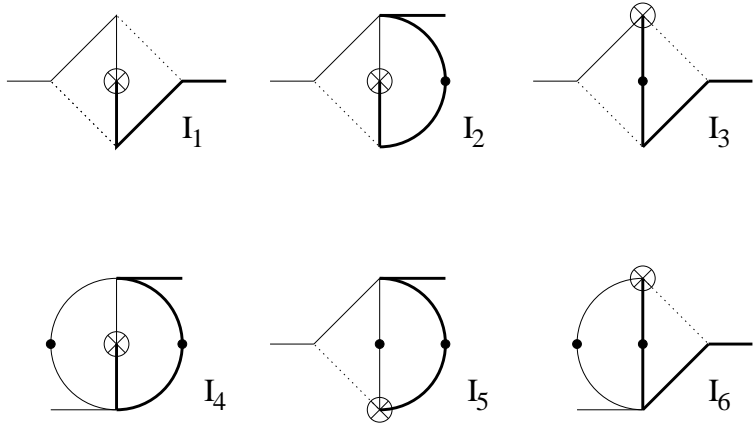,width=10cm}}
\caption[]{The scalar integrals $I_1 \ldots I_6$.
           The momentum $p_1$ enters from the left,
           $p_2-p_1$ enters at the vertex marked $\otimes$,
           $p_2$ leaves at the right. The thin (thick) lines
           symbolize quark propagators with mass $m_1$ ($m_2$).
           The dotted lines denote massless propagators. A line
           with a heavy dot on it means the corresponding
           propagator is squared.}
\label{fig:ddm}
\end{figure}

Let us take the six-propagator integral
\begin{equation}
I_1(m_1,m_2) \; = \; \int \int \mbox{d}^d k  \; \mbox{d}^d l \;
\frac{1}{P_1 P_2 P_3 P_6 P_7 P_8}
\end{equation}
shown in figure~\ref{fig:ddm} as an example. It is one of the contributions
that come from diagram $c_2$ in figure~\ref{fig:diagrams}.
This particular integral has no ultraviolet divergence, but it does
contain an infrared divergence coming from the region where both
$k$ and $l$ are small. Our goal is to express it in terms of
other integrals in which one (or both) of the massless propagators $P_7$
and $P_8$ are cancelled. Let $p_{1,2}=m_{1,2}q$ where $q$ is a
fixed four-vector with $q^2=1$. First, consider differentiation
with respect to $m_2$. Using the identities,
\begin{eqnarray}
\frac{\partial}{\partial m_2} \left( \frac{m_2}{P_6} \right)
& = &\frac{P_7}{P_6^2} \\
\frac{\partial}{\partial m_2} \left( \frac{m_1-m_2}{P_2} \right)
& = & - \frac{P_1}{P_2^2} \, ,
\end{eqnarray}
we get
\begin{equation}
\frac{\partial}{\partial m_2} \left\{
 (m_1-m_2) m_2 I_1(m_1,m_2) \right\}
=
(m_1-m_2)\, I_2(m_1,m_2) - m_2 I_3(m_1,m_2) \, ,
\end{equation}
where
\begin{equation}
I_2(m_1,m_2) \; = \; \int \int \mbox{d}^d k  \; \mbox{d}^d l \;
\frac{1}{P_1 P_2 P_3 P_6^2 P_8}  \, ,
\end{equation}
\begin{equation}
I_3(m_1,m_2) \; = \; \int \int \mbox{d}^d k  \; \mbox{d}^d l \;
\frac{1}{P_2^2 P_3 P_6 P_7 P_8}  \, .
\end{equation}
The integral $I_2$ is already finite, but it can be simplified
further by repeating the above procedure with $m_1$ instead of
$m_2$:
\begin{equation}
\frac{\partial}{\partial m_1} \left\{
 (m_2-m_1) m_1 I_2(m_1,m_2) \right\}
=
(m_2-m_1)\, I_4(m_1,m_2) - m_1 I_5(m_1,m_2) \, ,
\end{equation}
\begin{equation}
I_4(m_1,m_2) \; = \; \int \int \mbox{d}^d k  \; \mbox{d}^d l \;
\frac{1}{P_1 P_2 P_3^2 P_6^2}  \, ,
\end{equation}
\begin{equation}
I_5(m_1,m_2) \; = \; \int \int \mbox{d}^d k  \; \mbox{d}^d l \;
\frac{1}{P_1^2 P_3 P_6^2 P_8}  \, .
\end{equation}
On the other hand, $I_3$ is still just as divergent as $I_1$. However,
since it no longer contains $P_1$, it is now easy to cancel $P_8$:
\begin{equation}
\frac{\partial}{\partial m_1} \left\{
 m_1 I_3(m_1,m_2) \right\}
=
 I_6(m_1,m_2) =  \; \int \int \mbox{d}^d k  \; \mbox{d}^d l \;
\frac{1}{P_2^2 P_3^2 P_6 P_7}  \, .
\end{equation}
From figure~\ref{fig:ddm}, it is obvious that
$I_6(m_1,m_2)=I_5(m_2,m_1)$. Thus, in order to obtain the integral
$I_1$, all one needs to know are the two finite, four-propagator
integrals $I_4$ and $I_5$, which are relatively easy to do
by Feynman parametrization, and, as an integration constant, the
value of $I_3(m,m)$, which can be calculated by recurrence relations.
Neglecting terms of order $\epsilon$, we find
\begin{eqnarray}
I_3(m,m) & = & \pi^d \Gamma^2(1+\epsilon) \frac{m^{-4-4\epsilon}}{4}
\left\{ \frac{1}{\epsilon} + 3 \zeta(2) - 4 \right\} ,
\\
I_4(m_1,m_2)  & = & \frac{\pi^4}{8 m_1^2 m_2^2}
\left\{ 3 \zeta(2) - \frac{2}{u} \left( \Li{2}{u} - \Li{2}{-u} \right)
+ \frac{u^2-1}{u} \log \left(\frac{1+u}{1-u}\right)
\right\} ,
\\
I_5(m_1,m_2)  & = & \frac{\pi^4}{8 m_1^2 m_2^2}
\left\{ 2 - \frac{2}{u} - 3 \zeta(2) + 2 \Li{2}{u} - 2 \Li{2}{-u}
+ \frac{1-u^2}{u^2} \log \left(\frac{1+u}{1-u}\right)
\right\} ,
\end{eqnarray}
where the dimensionless variable $u$ is defined as $(m_1-m_2)/(m_1+m_2)$,
and, after performing the final integrations over $m_1$ and $m_2$,
\begin{equation}
I_1(m_1,m_2) = \pi^d \Gamma^2(1+\epsilon)
\frac{m_1^{-2\epsilon} m_2^{-2\epsilon}}{4 m_1^2 m_2^2}
\left\{ \frac{1}{\epsilon} + 3 \zeta(2) - 12
       + \frac{2}{u} \left( \Li{2}{u} - \Li{2}{-u} +
       \log \left(\frac{1+u}{1-u}\right) \right)
\right\} .
\end{equation}
It is comforting to see that the result is symmetric in $m_1$ and
$m_2$, as it clearly should be.

We will not present formulae for all the non-factorizable
two-mass scalar integrals (\ref{eq:generalscalarint}) that
appear in the diagrams of figure~\ref{fig:diagrams}, but mention
that they can all be expressed in terms of the following basic
set of logarithms and polylogarithms:
\begin{equation}
\label{eq:basicpolylogs}
\begin{array}{c}
 \log \left(\frac{1+u}{1-u}\right)\,,\quad
 \log \left(1+u\right)\,,\quad
 \Li{2}{u}\,,\quad
 \Li{2}{-u}\,,\quad
 \Li{2}{\frac{2 u}{u+1}}\,,
\\[1ex]
 \Li{3}{\frac{u}{u+1}}\,,\quad
 \Li{3}{\frac{u}{u-1}}\,,\quad
 \Li{3}{\frac{2 u}{u+1}}\,,\quad
 \Li{3}{\frac{2 u}{u-1}}\,,\quad
 \Li{3}{\frac{4 u}{(u+1)^2}}\,,\quad
 \Li{3}{\frac{-4 u}{(u-1)^2}}\,.
\end{array}
\end{equation}
These functions are all real and analytic in the range
$-1<u<1$, which corresponds to real, positive masses $m_1$, $m_2$.
It is easy to verify that their only singularities are located
at $u=\pm 1$, and in some cases $u=\infty$, in accordance
with the general properties inferred above from the Landau
equations.
The functions that have branch points at $u=\infty$ only
occur in diagrams with massive three-particle cuts, and
the trilogarithms only occur in five-propagator integrals.

We applied various checks on the results for the scalar integrals.
First of all, we evaluated the ones that are finite in $d=4$
by direct numerical integration in momentum space, basically
by a straightforward extension of the method originally
proposed in \cite{Kreimer2L2P} for the two-point two-loop ``master" diagram
with general masses. However, there are a few finite integrals that
cannot easily be evaluated by this method, unless an important
modification is made, which is explained in the appendix.
Then, there are a number of ``sunset" integrals, which we checked
numerically by means of dispersion relations \cite{PP},
and a few other two-point functions we compared with the
program package Xloops~\cite{XLoops}.

A simple analytical check that can be applied to all two-mass integrals
is to compare the first few terms of their Taylor expansions in
$u$ with what one gets by expanding the propagators around
the equal mass point ($m_1=m_2$) under the integral sign (as
was done from the outset in \cite{Czarnecki}). While
this is a powerful test, it is not sensitive to possible
mistakes in the integration constants in diagrams calculated
by differentiating and then reintegrating with respect to
the masses. Therefore, for some integrals, we also looked
at their asymptotic behaviour in the limit when $m_1$ or $m_2$
vanishes ($u\to \mp 1$). In this limit, the diagrams are again
reduced to one-scale integrals, but it is more complicated than the
equal mass limit because, in general, the diagrams become more
strongly divergent when one of the masses vanishes. Nevertheless,
it is possible to obtain the correct asymptotic behaviour by
expanding under the integral sign, provided one adds certain
counterterms corresponding to the Taylor expansion of certain
subdiagrams of the diagram considered, following the prescription
for asymptotic expansions on the mass shell given in
\cite{SmirnovMassShell}.

To conclude this list of consistency checks, we mention two tests that
we applied to the contributions of complete diagrams, rather than
individual scalar integrals. We verified that if
$m_1$ and $m_2$ are interchanged, diagrams $a_1$ and $a_2$ in
figure~\ref{fig:diagrams} are transformed into each other,
and similarly
$b_1 \leftrightarrow b_3$,
$b_2 \leftrightarrow b_2$,
$c_1 \leftrightarrow c_3$,
$c_2 \leftrightarrow c_2$,
$d   \leftrightarrow d  $,
$e   \leftrightarrow e  $ and
$f   \leftrightarrow f  $. This is a consequence of
charge conjugation symmetry. Secondly, by putting the
masses of the two external quarks equal to each other,
we reproduced the two-loop on-shell quark wave-function
renormalization constants in \cite{BGS}. Except for
one part --- the
contribution\footnote{Equation (22) in \cite{BGS}}
of a quark loop whose mass is neither zero, nor equal to that of the
external quark line --- which we calculated separately,
we could obtain all the wave-function renormalization factors by
simply substituting $u=0$ in our formulae for contributions
to $\eta_V$.
\section{Results}
\label{sec:res}
Here, we present our results for $\eta_{V,A}$ up to ${\cal O}(\alpha_s^2)$
in QCD with $N$ colours and $N_L$ flavours of massless quarks, but no
$t$ quarks. The colour factors are given by
\begin{equation}
 C_A=N,
\hspace{5ex}
 C_F=\frac{N^2-1}{2N} \, ,
\hspace{5ex}
 T_F=\frac{1}{2} \, .
\end{equation}
Following \cite{Czarnecki,CM}, we renormalize the
strong coupling constant in the $\overline{MS}$ scheme at the scale
$M=\sqrt{m_b m_c}$ but use on-shell renormalization for the masses.
This renormalization procedure has the nice property that it respects
the symmetry under $m_b \leftrightarrow m_c$ mentioned at the end of
the previous section.
Therefore, $\eta_{V,A}$ are even functions of the variable:
\begin{equation}
u = \frac{m_b-m_c}{m_b+m_c} \, .
\end{equation}
This symmetry was exploited in \cite{Czarnecki} by using
the variable
$\rho=(m_b-m_c)^2/(m_b m_c)=4u^2/(1-u^2)$
rather than
$\delta=(m_b-m_c)/m_b=2u/(1+u)$
as an expansion parameter, in order to obtain a more rapidly
converging series.
%
%
%
%

It turns out that the trilogarithms
$\Li{3}{\frac{u}{u+1}}$
and
$\Li{3}{\frac{u}{u-1}}$,
which appear in the contributions of diagrams $c_1$, $c_2$ and $c_3$,
cancel out in the sum $c_1+c_2+c_3$.
The other functions (\ref{eq:basicpolylogs}) appear in the following
combinations:
\begin{eqnarray}
\ell       & = & \log \left(\frac{1+u}{1-u}\right) \\
{\cal L}_1 & = & \Li{2}{u} - \Li{2}{-u}        \\
{\cal L}_2 & = & \Li{2}{\frac{2 u}{u+1}} + \frac{1}{4} \ell^2  \\
{\cal L}_3 & = & \Li{3}{\frac{2 u}{u+1}} + \Li{3}{\frac{2 u}{u-1}}
        + \frac{1}{6} \ell^3 + \frac{2}{3} \ell\ \Li{2}{\frac{2 u}{u+1}} \\
{\cal L}_4  & = & \Li{3}{\frac{4 u}{(u+1)^2}} + \Li{3}{\frac{-4 u}{(u-1)^2}}
        + \frac{4}{3} \ell^3 +\frac{16}{3} \ell\ \Li{2}{\frac{2 u}{u+1}}
\nonumber \\ &&
          - \zeta(2)\ \left(4 \log (1+u) -2 \ell\right)
          - \frac{8}{3} \ell\ \left(\Li{2}{u} - \Li{2}{-u}\right)
\, .
\end{eqnarray}
Note that $\ell$, ${\cal L}_1$ and ${\cal L}_2$ are odd functions
of $u$, while ${\cal L}_3$ and ${\cal L}_4$ are even. The abbreviations
$a$ and $z_3$ are defined by
\begin{eqnarray}
a   & = & \frac{\alpha_s(\sqrt{m_b m_c})}{4\pi} \\
z_3 & = &  \zeta(3) - 4 \log(2)\ \zeta(2) \, .
\end{eqnarray}
Finally, the results are
\begin{eqnarray}
\label{eq:etav}
\eta_V = && 1
  + a C_F \left( \ell\  \frac{3}{u}
              - 6
            \right)
\nonumber \\ &&
  + a^2 \left[
    C_F T_F N_L \left\{
    \ell\  \frac{(  - \frac{2}{3} )}{u} + \frac{4}{3}
             \right\}
\;\;\;\; \right. \nonumber \\ &&
  + C_F T_F \left\{
    \zeta(2)\  \frac{(  - 104 u^2 + 464 u^4 + 1144 u^6 + 32 u^8 )}{(u^2-1)^4}
    + \ell\ \frac{(  - \frac{64}{3} + \frac{200}{3} u^2 + \frac{344}{3} u^4 )}
                  {u (u^2-1)^2}
\right. \nonumber \\ &&
\;\;\;\;
    + \ell^2\  \frac{(  - 16 + 224 u^2 + 896 u^4 + 416 u^6 + 16 u^8 )}
                   {(u^2-1)^4}
\nonumber \\ &&
\;\;\;\;
    + {\cal L}_1 \frac{(  - 32 + 128 u^2 + 1616 
         u^4 + 1312 u^6 + 48 u^8 )}{u (u^2-1)^4}
\nonumber \\ &&  \left.
\;\;\;\;
    + {\cal L}_2  \frac{ 32 - 128 u^2 - 2368 u^4
          - 3200 u^6 - 480 u^8 }{u (u^2-1)^4}
    + \frac{ \frac{128}{3} + \frac{296}{3} u^2 + \frac{344}{3} u^4 }
           {(u^2-1)^2}
          \right\}
\nonumber \\ &&
  + C_F^2 \left\{
    \zeta(2)\ \ell\  \frac{(  - 16 )}{u}
    + \zeta(2)\  \frac{(  - 32 + 48 u^2 )}{u^2-1}
    + \ell\  \frac{(  - \frac{89}{6} )}{u}
    + \ell^2\  \frac{ 3 - \frac{9}{2} u^2 + \frac{11}{2} u^4}{u^2 (u^2-1)}
\right. \nonumber \\ && \left.
\;\;\;\;
    + {\cal L}_2  \frac{(  - 16 )}{u (u^2-1)}
    + {\cal L}_3  \frac{ 24 - 96 u^2 }{u^2}
    +  \frac{53}{3}
         \right\}
\nonumber \\ &&
  + C_F(C_A-2C_F) \left\{
    \zeta(2)\ \ell\  \frac{(  - 8 u )}{u^2-1}
    + \zeta(2)\  \frac{ 10 u^2 }{u^2-1}
    + \ell\  \frac{ \frac{17}{6} }{u}
    + \ell^2\  \frac{ 2 + 2 u^2 }{u^2-1}
\right. \nonumber \\ && \left. \left.
\;\;\;\;
    + {\cal L}_1 \frac{ 4 u }{u^2-1}
    + {\cal L}_2  \frac{(  - 16 u )}{u^2-1}
    - 24 {\cal L}_3
    + {\cal L}_4  \frac{ 12 - 6 u^2 }{u^2-1}
    + z_3  \frac{(  - 6 u^2 )}{u^2-1}
    - \frac{17}{3}
        \right\} \right]
\end{eqnarray}
and
\begin{eqnarray}
\label{eq:etaa}
\eta_A = && 1
  + a C_F \left(
    \ell\  \frac{3}{u}
     - 8 
            \right)
\nonumber \\ &&
  + a^2 \left[
    C_F T_F N_L \left\{
    \ell\  \frac{(  - \frac{10}{3} )}{u}
    + \frac{88}{9}
             \right\}
\right. \nonumber \\ &&
  + C_F T_F \left\{
    \zeta(2)\  \frac{(  - 64 - \frac{88}{3} u^2 + \frac{2512}{3} u^4 + 
         \frac{2312}{3} u^6 + \frac{64}{3} u^8 )}{(u^2-1)^4}
\right. \nonumber \\ &&
\;\;\;\;
    + \ell\  \frac{(  - \frac{16}{3} + 120 u^2 + \frac{136}{3}  u^4 )}
                  {u (u^2-1)^2}
    + \ell^2\  \frac{ \frac{64}{3} + \frac{1120}{3} u^2 + \frac{2464}{3} u^4
          + \frac{928}{3} u^6 + \frac{32}{3} u^8 }{(u^2-1)^4}
\nonumber \\ &&
\;\;\;\;
    + {\cal L}_1 \frac{(  - \frac{32}{3}
     + \frac{1472}{3} u^2 + 
         \frac{4528}{3} u^4 + \frac{3104}{3} u^6 + 48 u^8 )}{u (u^2-1)^4}
\nonumber \\ && \left.
\;\;\;\;
    + {\cal L}_2  \frac{ \frac{32}{3} - \frac{1664}{3} u^2
    - \frac{8000}{3} u^4 - \frac{7808}{3} u^6 - \frac{992}{3} u^8 }
   {u (u^2-1)^4}
    + \frac{ \frac{1016}{9} + \frac{584}{9} u^2 + \frac{704}{9} u^4 }
           {(u^2-1)^2}
          \right\}
\nonumber \\ &&
  + C_F^2 \left\{
    \zeta(2)\ \ell\  \frac{(  - 16 )}{u}
    + \zeta(2)\  \frac{(  - 48 + \frac{160}{3} u^2 )}{u^2-1}
    + \ell\  \frac{(  - \frac{53}{6} )}{u}
    + \ell^2\  \frac{(  - \frac{4}{3} - \frac{5}{2} u^2 + \frac{31}{6} 
         u^4 )}{u^2 (u^2-1)}
\right. \nonumber \\ && \left.
\;\;\;\;
    + {\cal L}_2  \frac{ \frac{16}{3} - \frac{32}{3} u^2 }{u (u^2-1)}
    + {\cal L}_3  \frac{(  - 8 - 64 u^2 )}{u^2}
    - \frac{190}{9}
         \right\}
\nonumber \\ &&
  + C_F(C_A-2C_F) \left\{
    \zeta(2)\ \ell\  \frac{ \frac{32}{3} - \frac{56}{3} u^2 }{u (u^2-1)}
    + \zeta(2)\ \frac{ \frac{40}{3} - \frac{46}{3} u^2 + \frac{70}{3} u^4 }
                     {(u^2-1)^2}
    + \ell\  \frac{ \frac{61}{6} }{u}
\right. \nonumber \\ &&
\;\;\;\;
    + \ell^2\  \frac{(  - \frac{2}{3} + \frac{28}{3} u^2 + \frac{26}{3} 
         u^4 + 4 u^6 )}{u^2 (u^2-1)^2}
    + {\cal L}_1
       \frac{ \frac{32}{3} + \frac{52}{3} u^2 + \frac{44}{3} 
         u^4 }{u (u^2-1)^2}
\nonumber \\ && \left. \left.
\;\;\;\;
    + {\cal L}_2  \frac{(  - \frac{32}{3} - \frac{112}{3} u^2 - \frac{112}{
         3} u^4 )}{u (u^2-1)^2}
    - 24 {\cal L}_3
    + {\cal L}_4  \frac{ 4 + 2 u^2 }{u^2-1}
    + z_3  \frac{ 4 - 10 u^2 }{u^2-1}
    - \frac{302}{9}
        \right\} \right] \, .
\end{eqnarray}
Note that every single term in the expressions (\ref{eq:etav})
and (\ref{eq:etaa}) is manifestly analytic in $u$ for at least
$|u|<1$. This fact, combined with the symmetry
$u\leftrightarrow-u$, proves (once more) that $\eta_{V,A}$
can be represented by convergent power series in $u^2$
for all finite positive values of $m_b$ and $m_c$.
\section{Conclusion}
\label{sec:conc}
We have performed an independent calculation of the vector
and axial vector form factors $\eta_V$ and $\eta_A$ that
describe the decay of a $b$ quark into a $c$ quark at zero recoil
up to order $\alpha_s^2$. We have compared our expressions
(\ref{eq:etav}) and (\ref{eq:etaa}) with the corresponding formulae
in \cite{CM} by rewriting the latter in terms of the
functions~(\ref{eq:basicpolylogs}) used in this paper, and
found that the two calculations are in perfect agreement.
Both confirm the series expansion in the mass difference $m_b-m_c$
obtained earlier in \cite{Czarnecki}.

The actual numerical values of $m_b$ and $m_c$ are such that
a few terms of the series expansion are already sufficient to
achieve the accuracy that is needed in practice ($u^2\approx 0.29$).
On the other hand, the exact analytical formulae are not only
valid for $b \to c$ decays, but also for other cases, such as
$t \to b$, where the mass ratio is larger, and the series expansion
converges more slowly. They also allow one to study the limit
when one of the quark masses goes to zero, which is impossible
if only a limited number of terms of an expansion in the mass
difference are known.

In the appendix, we have identified the cause of a numerical
difficulty that arises when the two-dimensional numerical
integration method of \cite{Kreimer2L2P} is applied to
certain two-loop diagrams with several coinciding thresholds,
and suggested an alternative method that solves the problem.

We would like to thank J.G.~K{\"o}rner for suggesting this project
and P. Post for helping us check some sunset diagrams. We are also
grateful to A.~Czarnecki, A.G.~Grozin, K.~Melnikov, M.~Neubert
and D.~Pirjol for discussions. This work was partly supported by
the Graduiertenkolleg
``Elementarteilchenphysik bei mittleren und hohen Energien'' at the
Johannes Gutenberg University in Mainz.
\section*{Appendix}
In this appendix, some details of the numerical integration
methods we used to check the scalar two-loop integrals appearing
in this work are given. We restrict ourselves to cases that are
both ultraviolet and infrared finite in $d=4$. Basically, we
follow the approach proposed in \cite{Kreimer2L2P} for the
numerical integration of the scalar two-point two-loop
master diagram with general masses. The fact that the
diagrams occurring here are three-point diagrams is not
a problem: since the external momenta $p_1$ and $p_2$
are proportional to each other, they have a common rest frame, and all
the steps in the derivation of the integral representation
given in \cite{Kreimer2L2P} can be repeated with only
a few trivial changes in the formulae. However, the fact
that $p_1$ and $p_2$ are on their respective mass shells
does give rise to some difficulties, which we discuss here.

\begin{figure}[htb]
\centerline{\epsfig{file=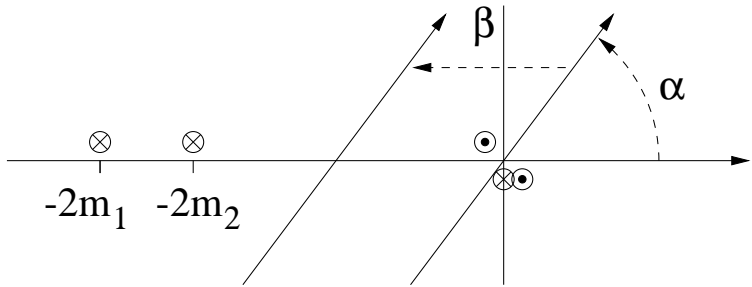,width=10cm}}
\caption[]{Singularities in the complex $k_0$-plane
           originating from the quark propagators
           $P_5$ and $P_6$ ($\otimes$), and from
           the gluon propagator $P_7$ ($\odot$).
           While the integration contour can always be rotated
           away from the real axis by an angle $\alpha$, it can
           only be shifted away from the origin in diagrams not
           containing the gluon propagator $P_7$.
           }
\label{fig:k0plane}
\end{figure}

By writing the loop momenta $k$ and $l$ in the common rest frame
of $p_1$ and $p_2$ as $(k_0,\vec{k}_{\perp})$, $(l_0,\vec{l}_{\perp})$,
and integrating out their space components
$\vec{k}_{\perp}$, $\vec{l}_{\perp}$,
we obtain representations of the form
\begin{equation}
\label{eq:intk0l0}
\int_{-\infty}^{\infty}\int_{-\infty}^{\infty}
\mbox{d} k_0\;\mbox{d} l_0 \;
F(k_0,l_0;\rho)
\end{equation}
for the scalar integrals (\ref{eq:generalscalarint}). Here,
we have explicitly indicated the dependence on the small
imaginary quantity $i\rho$ which is added to all propagator
denominators in accordance with the causal prescription.
As a function of the complex variable $k_0$, the integrand
$F(k_0,l_0;\rho)$
has singularities (branch points) located at the positions
shown in figure~\ref{fig:k0plane}. Identical pictures can
be drawn for the complex variable $l_0$ and for the
combination $k_0+l_0$. When the limit $\rho\to 0$ is taken,
the singularities move onto the real $k_0$ and $l_0$ axes.
In general, this makes the integrals along the real axes
more difficult to evaluate numerically and in some cases
even divergent. Sometimes, for example when the scalar
integral under consideration does not contain either of the gluon
propagators $P_7$ and $P_9$, the problem can be solved by
rotating the integration contours away from the real axes
by an angle $\alpha$ (which must be the same for $k_0$ and $l_0$),
$l_0 \to e^{i\alpha} l_0$,
$k_0 \to e^{i\alpha} k_0$,
and subsequently shifting (one of) them away from the origin,
$e^{i\alpha} k_0 \to e^{i\alpha} k_0 - \beta$, as shown
in figure~\ref{fig:k0plane}. Scalar integrals we checked in this way
include $I_2$ in figure~\ref{fig:ddm} and
$\int \int \mbox{d}^4 k \, \mbox{d}^4 l / (P_1 P_3 P_6^3 P_8)$.

This escape is not possible in diagrams where both
the $k_0$ and $l_0$ contours are trapped at the origin by
gluon propagators, such as, e.g.,
\begin{equation}
\label{eq:nasty}
\int \int \mbox{d}^4 k \; \mbox{d}^4 l \;
\frac{1}{P_1 P_3 P_6 P_7 P_8} \, .
\end{equation}
We stress that if the limit $\rho\to 0$ is taken
after all integrations have been performed, one
obtains a finite result for the integral
(\ref{eq:nasty})\footnote{
In the case of equal masses $m_1=m_2$, it reduces to
the one difficult one-scale integral
$N(1,1,1,1,1)=6\,\zeta(2) \log(2) - \frac{3}{2}\,\zeta(3)$ needed
in two-loop QCD or QED corrections to on-shell
fermion propagators \cite{Broadhurst}.}. However, one
would like to set $\rho$ to zero, or at least, very close
to zero, {\em before} performing the $k_0$ and $l_0$ integrations
in (\ref{eq:intk0l0}), and, if one attempts to do so, a non-integrable
singularity appears at the point $k_0=l_0=0$. More specifically,
$F(\lambda k_0,\lambda l_0;\rho=0) \sim 1/\lambda^2$ as $\lambda\to 0$.
In the original integral (\ref{eq:nasty}), this logarithmic divergence comes
from the region where $\vec{k}_\perp$ and $\vec{l}_\perp$ tend to
zero, while $k_0$ and $l_0$ are of order $\vec{k}_\perp^2$,
$\vec{l}_\perp^2$, as one can see by
rescaling
\begin{equation}
\vec{k}_\perp \to \lambda \vec{k}_\perp \, ,
\hspace{3em}
\vec{l}_\perp \to \lambda \vec{l}_\perp \, ,
\hspace{3em}
k_0     \to \lambda^2 k_0 \, ,
\hspace{3em}
l_0     \to \lambda^2 l_0 \, .
\end{equation}
Under this transformation, the integration measure in
(\ref{eq:nasty}) scales like $\lambda^{10}$, while,
for small $\lambda$, the integrand
goes like $1/\lambda^{10}$.

The problem can be solved very easily by interchanging the order
of the integrations. That is, we first perform the $k_0$, $l_0$
and all angular integrations analytically, leaving
two integrations over $k_{\perp}=|\vec{k}_{\perp}|$ and
$l_{\perp}=|\vec{l}_{\perp}|$ to be done numerically.

Below, we shall give an explicit formula for the resulting
integrand. Although our main reason for deriving it was the need
for an independent check on our analytic results for (\ref{eq:nasty})
and a few other, equally nasty cases, the formula is valid for
a much more general diagram:
\begin{equation}
\label{eq:defefirst}
J = \int \int \mbox{d}^4 k  \; \mbox{d}^4 l \;
\frac{1}{D_1 D_2 D_3 D_4 D_5}
 \, ,
\end{equation}
where
\begin{eqnarray}
&& D_1 = (k+p_1)^2-m_1^2+i\rho \hspace{5em}
   D_4 = (l+p_4)^2-m_4^2+i\rho \nonumber \\
&& D_2 = (k+p_2)^2-m_2^2+i\rho \hspace{5em}
   D_5 = (l+p_5)^2-m_5^2+i\rho           \\
&& D_3 = (k+l+p_3)^2-m_3^2+i\rho \nonumber
\, .
\end{eqnarray}
with the restriction that all momenta $p_i$ are proportional
to one another. Apart from that, the $p_i$ and the masses $m_i$
are arbitrary. Depending on the choice of the $p_i$, (\ref{eq:defefirst})
then corresponds to a diagram with two, three or four external legs.
We use this notation for the sake of flexibility, even though it has
some redundancy (which could be used, for example, to set $p_2=p_5=0$).
Working in the rest frame of the momenta $p_i$, we find the following
representation:
\begin{eqnarray}
\label{eq:efirst}
J & = & 4 \pi^4 \int_0^{\infty}\mbox{d} k_{\perp}
            \int_0^{\infty}\mbox{d} l_{\perp}
k_{\perp} l_{\perp} \left\{
\frac{1}{u_1 u_4}
\left( \frac{F_3^+(u_1-p^0_1+u_4-p^0_4)}{C_{12}^+ C_{45}^+}
   +  \frac{F_3^-(-u_1-p^0_1-u_4-p^0_4)}{C_{12}^- C_{45}^-}
\right) \right.
\nonumber \\[0.2cm]
& & \hphantom{4 \pi^4 \int_0^{\infty}\mbox{d} k_{\perp}
            \int_0^{\infty}\mbox{d} l_{\perp}
           k_{\perp} l_{\perp}\left\{\right.}
\left.
\vphantom{\frac{F_3^+(u_1-p^0_1+u_4-p^0_4)}{C_{12}^+ C_{45}^+}}
+ (1 \leftrightarrow 2) + (4 \leftrightarrow 5)
+ (1 \leftrightarrow 2, 4 \leftrightarrow 5)
\vphantom{\frac{1}{u_1 u_4}}
\right\} \, ,
\end{eqnarray}
with
\begin{equation}
F_3^{\pm}(x) = \log \left( \frac{x+p^0_3\pm u_3^+}
                                {x+p^0_3\pm u_3^-} \right)
\, ,
\end{equation}
\begin{equation}
\label{eq:Cijdef}
C_{ij}^{\pm} = {(\pm u_i - p^0_i + p^0_j)}^2 - u_j^2
\, ,
\end{equation}
and
\begin{eqnarray}
&& u_1  =  \sqrt{m_1^2+k_{\perp}^2-i\rho} \hspace{5em}
   u_4  =  \sqrt{m_4^2+l_{\perp}^2-i\rho} \nonumber \\
&& u_2  =  \sqrt{m_2^2+k_{\perp}^2-i\rho} \hspace{5em}
   u_5  =  \sqrt{m_5^2+l_{\perp}^2-i\rho}           \\
&& u_3^{\pm} =  \sqrt{m_3^2+(k_{\perp}\pm l_{\perp})^2-i\rho} \,.
\nonumber
\end{eqnarray}
The representation (\ref{eq:efirst}) can be applied directly as it stands
to the case (\ref{eq:nasty}). In general though, there will be poles
near the real axes at the points where the $C_{ij}^{\pm}$ vanish. They
can easily be avoided by a rotation of the contours:
$l_{\perp} \to e^{-i\alpha} l_{\perp}$,
$k_{\perp} \to e^{-i\alpha} k_{\perp}$.

\end{document}